\documentclass[aps,prl,11pt,preprint]{revtex4}
\usepackage{graphics,dcolumn}
\usepackage{graphicx}
\usepackage{amsmath}

\begin{document}

\title{A Quantum-Mechanics Molecular-Mechanics scheme for extended systems}
\author{Diego Hunt$^{\dag}$}
\author{Veronica M. Sanchez$^{\ddag}$}
\author{Dami\'an A. Scherlis$^{\dag,*}$}

\affiliation{$^\dag$Departamento de Qu\'imica Inorg\'anica, Anal\'itica
y Qu\'imica F\'isica/INQUIMAE, Facultad de Ciencias Exactas
y Naturales, Universidad de Buenos Aires, Ciudad Universitaria,
Pab. II, Buenos Aires (C1428EHA) Argentina}

\affiliation{$^\ddag$Centro de Simulaci\'on Computacional Para Aplicaciones Tecnol\'ogicas,
Polo Cient\'ifico Tecnol\'ogico, CONICET, Godoy Cruz 2201, Buenos Aires, Argentina}


\begin{abstract}

We introduce and discuss a hybrid quantum-mechanics molecular-mechanics (QM-MM) approach 
for Car-Parrinello DFT simulations with pseudopotentials and planewaves basis, designed for
the treatment of periodic systems.
In this implementation the MM atoms are considered as additional QM ions having fractional
charges of either sign,
which provides conceptual and computational simplicity
by exploiting the machinery already existing in planewave codes to deal with electrostatics
in periodic boundary conditions.
With this strategy, both the QM and MM regions are contained in the same supercell, which
determines the periodicity for the whole system.
Thus, while this method is not meant to compete with  non-periodic QM-MM schemes able
to handle extremely large but finite MM regions,
it is shown that for periodic systems of a few hundred atoms,
our approach provides substantial savings in computational times
by treating classically a fraction of the particles.
The performance and accuracy of the method is assessed through the study
of energetic, structural, and dynamical
aspects of the water dimer and of the aqueous bulk phase. Finally, the QM-MM scheme is
applied to the computation of the vibrational spectra of water layers adsorbed at the TiO$_2$ anatase (101)
solid-liquid interface. This investigation suggests that the inclusion of a second monolayer
of H$_2$O molecules is sufficient to induce on the first adsorbed layer, a vibrational dynamics
similar to that taking place in the presence of an aqueous environment.
The present QM-MM scheme appears as a very interesting tool to
efficiently perform molecular dynamics simulations of complex condensed matter systems,
from solutions to nanoconfined fluids to different kind of interfaces.

\end{abstract}

\date{\today}
\maketitle

\section{Introduction}

In the context of molecular simulations, hybrid Quantum Mechanics-Molecular Mechanics (QM-MM)
schemes consider the system as a sum of two parts: solute (QM fragment) and
solvent (MM fragment) \cite{monard, luque, gao1997, eichinger, elola1999,yarne2001, Laio2002,
crespo2003,mao-hua2003, laino2005, laino2006,vms2006,jctc_3_628,rinaldo2007,nano2,sushko2010,bongards2010,
jp8071712, jcp_139_244108, jctc_9_5086}.
The particles are assigned to one of these two groups according to their role:
atoms directly involved in bonds breaking or forming, or in polarization or charge
transfer effects, must be considered in the QM region, whereas those atoms not participating
in these processes are included within the MM subsystem.
These two groups are described at different levels with different Hamiltonians, but they interact
with each other, generally self-consistently.

QM-MM schemes have been applied extensively and successfuly along the last couple of decades,
to model finite chemical and biological systems.
The impact of this methodology has been acknowledged through the
Chemistry Nobel Prize of the year 2013, which was awarded to some of its founders
for the development of multiscale modelling.
One of the major successes of this approach was in the study of chemical reactions inside the active
site of proteins. In this type of simulations the solute described quantum-mechanically
comprises the active site, while the rest of the protein plus hydration water molecules
are treated classically \cite{monard, luque, capece2008,bongards2010}.
Leaving aside the applications in biochemistry, hybrid QM-MM methodologies in a non-periodic setting
have also been employed in various other contexts, as for example
proton transfer reactions in water clusters \cite{elola1999,Lin2005-2},
or in different kinds of materials which were modeled as finite structures: 
these works have addressed solid-liquid \cite{mao-hua2003}, metal-organic \cite{sushko2010}, 
and oxide interfaces \cite{ellis2003}.

On the other hand, the QM-MM methodology applied in periodic boundary conditions (PBC) for
both the QM and MM parts, has been rarely
reported in the literature. A few examples imposing periodicity to
the MM region only to model dilute solutions can be found for
semiempirical  or first-principles approaches \cite{gao1997,nano2}. 
Laino et al. \cite{laino2005,laino2006} developed a QM-MM 
method in periodic boundary conditions based on Gaussian basis sets and multigrids to treat
the long-range interactions, which was tested on
the simulation of surface defects present at the $\alpha$-quartz phase of silica.
Other periodic QM-MM implementations with Gaussian basis sets have been
proposed based on the reduction of the electron density to point
charges, after which the classical Ewald summation can be applied \cite{jctc_1_2, jcp_139_244108}.
Such strategy has been implemented for both semiempirical \cite{jctc_1_2} 
and ab-initio \cite{jcp_139_244108} Hamiltonians. In this line, Golze and co-workers elaborated
a method for the treatment of metallic interfaces, where
the interactions between the quantum-mechanical adsorbate and the classical substrate
are handled at the molecular mechanics level \cite{jctc_9_5086}.
To the best of our knowledge only Yarne et al. \cite{yarne2001} developed a hybrid QM-MM methodology
imposing PBC  to the whole system in a pseudopotentials planewaves (PPW) code \cite{kohanoff}.
In this case electrons were confined to a smaller unit cell 
inside the supercell needed to describe the whole system, and periodicity 
was limited to 1 or 2-D \cite{yarne2001}.

In the present article, we present a formulation for hybrid QM-MM calculations based on density
functional theory (DFT) in a PPW framework.
In particular, this scheme has been devised for the
Car-Parrinello method as implemented in the Quantum Espresso code \cite{QuantumEspresso}.
The goal is to have available a hybrid QM-MM methodology in PBC appropriate to
describe condensed matter in complex environments, as solid-liquid or liquid-liquid interfaces,
where the main interest or the ``chemistry'' involves one of the two phases---the QM part---under
the influence of the other---the MM part. This could be useful
in solid-water interfaces, where the solid and any adsorbed species can be
described quantum-mechanically, while the solution may be modeled using a classical force-field.
We illustrate this kind of application through the study of titania in contact with an
aqueous phase.
The opposite representation, in which the solid constitutes the MM part, might also be appealing
if the interest were in the properties of the other phase,
as it could be the case of nanoconfined molecules or fluids.
To the best of our knowledge, no other QM-MM model has been
based on the present strategy, which we believe is a very interesting one
for atomistic simulations of interfaces or nano-spaces with high accuracy at an affordable
computational cost.


\section{QM-MM method in the Pseudopotential Plane Wave framework}

\subsection{Partitioning of the total energy}

In the DFT-PPW approach used in the Quantum Espresso code, 
the QM energy  can be cast as \cite{kohanoff,marx_hutter}:

\begin{equation}
 E_{QM} =  T_e[\rho] + E_H[\rho] + E_{ii} + E^{loc}_{PS}[\rho] + E_{PS}^{nl} + E_{XC}[\rho] 
\label{eqmtot}
\end{equation}

On the right hand side of the above equation, from left to right, there is the
kinetic energy of the electrons, the Hartree energy, the ion-ion repulsion,
the local and non-local contributions to the pseudopotential energies, and the exchange-correlation functional. 
Here $\rho(\textbf{r})$ is the electron charge density.

In the context of QM-MM models, the Hamiltonian and the energy of the system are written as:

\begin{equation}
 \hat{H}_{tot} = \hat{H}_{QM} + \hat{H}_{MM} + \hat{H}_{QM-MM}
\label{hamiltonian}
\end{equation}

\begin{equation}
 E_{tot} = E_{QM} + E_{MM} + E_{QM-MM}
\label{energy-qmmm1}
\end{equation}

\noindent where H$_{QM-MM}$ (and the related energy E$_{QM-MM}$) is a coupling term describing the
interaction between the two regions of the system.
In the MM region, atoms are typically treated as point charges of charge $Z_I$ interacting with
each other through electrostatics, dispersive-repulsive and harmonic potentials,
so that the molecular mechanics energy $E_{MM}$ is the sum of three contributions \cite{Leach1999}:

\begin{equation}
 E_{MM} = E_{ele} + E_{LJ} + E_{bond}
\label{energymm}
\end{equation}

\noindent where $E_{ele}$, $E_{LJ}$ and $E_{bond}$ denote the electrostatic, the Lennard-Jones,
and the bonding energy respectively, the later of which models the intramolecular
degrees of freedom.  In turn, these terms are normally computed as:

\begin{equation}
E_{ele} = \dfrac{1}{2} \sum_{I=1} {\sum_{J=1,J \neq I}{\dfrac{Z_IZ_J}{|\textbf{R}_I-\textbf{R}_J|}}}
\end{equation}

\begin{equation}
 E_{LJ} = \sum_I \sum_J 4 \epsilon_{IJ} \left[ \left(\dfrac{\sigma_{IJ}}{|\textbf{R}_I - 
\textbf{R}_J|}\right)^{12} - \left(\dfrac{\sigma_{IJ}}{|\textbf{R}_I - \textbf{R}_J|}\right)^{6} \right]
\label{lennardjones}
\end{equation}

\begin{equation}
 E_{bond} = \sum_{bonds}{\dfrac{k_i}{2}(l_i-l_{i0})^2} + \sum_{angles}{\dfrac{a_i}{2}
(\theta_i-\theta_{i0})^2} + \sum_{dihedrals}{\dfrac{v_n}{2}(1+ \cos (n\omega -\gamma))} 
\label{ebond}
\end{equation}

In the second of these three equations
$\sigma_{IJ}$ and $\epsilon_{IJ}$ are the Lennard-Jones radius and interaction energy between
atoms $I$ and $J$. In the last expression, $k_i, a_i$, and $v_n$, represent force constants
for the harmonic potentials controlling bond lengths, angles and torsions, respectively.
Within the MM region we will consider only water molecules,
which internal degress of freedom are described through the O-H distances and
H-O-H angles, and therefore the third term in the last equation will not be present.

The $E_{QM-MM}$ contribution appearing in equation \ref{energy-qmmm1} can normally be
explicitely written as the sum of an electrostatic and a non-electrostatic term.
As it will be shown below, however, in the working formula implemented here the electrostatic contribution
to $E_{QM-MM}$ can not be written separately, because it is intertwined
with the total electrostatic energy.  

\subsection{The electrostatic energy in the PPW framework}

In the PPW method, the electrostatic contribution comes from the sum of the second, third, and fourth
terms on the right hand side of equation \ref{eqmtot}:

\[
 E_{es}[\rho] =  E_H[\rho] + E^{loc}_{PS}[\rho] + E_{ii}  \]
\begin{equation}
    = \dfrac{1}{2} \iint{ \dfrac{\rho(\textbf{r})\rho(\textbf{r'})}{|\textbf{r}-\textbf{r'}|} d\textbf{r}d\textbf{r'}} 
+ \sum_{s=1}^{N} {\sum_{l=1}^{P_s} {\int{\rho(\textbf{r}) v_{PS}^{loc,s} (|\textbf{r} - \textbf{R}_I|) d\textbf{r} }}} 
+ \dfrac{1}{2} \sum_{I=1}^{P}{\sum_{J=1,J \neq I}^{P}{\dfrac{Z_IZ_J}{|\textbf{R}_I-\textbf{R}_J|}} } 
\label{eq-electrost}
\end{equation}

\noindent  where  $s$ indicates the atomic
species, $N$ is the number of different atomic species, and $v_{PS}^{loc,s}$ is the local part
of the pseudopotential
for each species. $Z_I$ is the ionic charge of the nuclei (which amounts to the
atomic number minus the valence electrons) and \textbf{R}$_I$ their positions.
$P$ stands for the number of ions and $P_s$ for the number of ions corresponding to the atomic species $s$.
We adopt the same convention used in the computational code, in which the sign of the
electronic charge is taken as positive and the ionic charge as negative.

Due to the long-range decay of electrostatic interactions,
$E_H[\rho]$, $ E^{loc}_{PS}[\rho]$ and $E_{ii}$ diverge if they are calculated separately.
It turns out to be convenient to introduce a fictitious ionic charge, $\rho_\alpha(\textbf{r})$, 
which is added to the Hartree energy term
and substructed from the other two. In this way,
a total neutral charge density is defined, $\rho_T(\textbf{r}) = \rho_{\alpha}(\textbf{r}) + \rho(\textbf{r})$,
and the electrostatic energy can be rewritten as:

\[
E_{es}[\rho] = \dfrac{1}{2} \iint{ \dfrac{\rho_T(\textbf{r})\rho_T(\textbf{r'})}{|\textbf{r}-\textbf{r'}|} d\textbf{r}d\textbf{r'}} + \int{ \rho(\textbf{r}) \left( \sum_{s=1}^{N} {\sum_{l=1}^{P_s} { v_{PS}^{loc,s} (|\textbf{r} - \textbf{R}_I|) d\textbf{r} }} - \int{ \dfrac{\rho_{\alpha}(\textbf{r'})}{|\textbf{r}-\textbf{r'}|} } \right) d\textbf{r} }\]
\begin{equation}\label{tot-den-1}
  + \dfrac{1}{2} \left( \sum_{I=1}^{P}{\sum_{J=1,J \neq I}^{P}{\dfrac{Z_IZ_J}{|\textbf{R}_I-\textbf{R}_J|}} } - \iint{\dfrac{\rho_{\alpha}(\textbf{r})\rho_{\alpha}(\textbf{r'})}{|\textbf{r}-\textbf{r'}|} d\textbf{r}d\textbf{r'}} \right) 
\end{equation}
The Hartree and the local pseudopotential contributions can be expanded in Fourier space, and
the ion-repulsions treated with the Ewald method. In particular, if
the ionic charge is defined as a sum of Gaussian functions centered on every nuclei,
\begin{equation}\label{gausianas}
 \rho_\alpha(\textbf{r}) = - \dfrac{\eta^3}{\pi^{3/2}} \sum_{I=1}^{P} Z_I e^{-2\eta ^2 | \textbf{r} - \textbf{R}_I |^2},
\end{equation}
after some manipulation the electrostatic energy can be expressed as \cite{kohanoff}:
\[ E_{es}[\rho] = \dfrac{\Omega}{2} \sum_{\textbf{G}} \dfrac{4\pi}{G^2} \tilde{\rho}_T (\textbf{G}) \tilde{\rho}_T (-\textbf{G}) 
+ \Omega \sum_{\textbf{G}} \left[ \sum_{s=1}^{N} S_s(\textbf{G}) \tilde{v}_{PS}^{loc,s}(\textbf{G}) 
-\dfrac{4\pi}{G^2}\tilde{\rho}_\alpha(\textbf{G}) \right] \tilde{\rho} (-\textbf{G}) \]
\begin{equation}\label{elec-final}
+ \dfrac{1}{2} \sum_{I=1}^{P} \sum_{J \neq 1}^{P} Z_I Z_J \left[ \sum_{n=-n_{max}}^{n_{max}} \dfrac{erfc(|\textbf{R}_I + nL - \textbf{R}_J|\eta)}{|\textbf{R}_I + nL - \textbf{R}_J|} \right] 
 - \dfrac{\eta}{\sqrt{\pi}} \sum_{I=1}^{P} Z^2_I .
\end{equation} 
where $\Omega$ is the volume of the supercell, $1/\eta$ is a cutoff distance parameter, 
and $S_s(\textbf{G})$ is
an atomic structure factor for each species $s$,
\begin{equation}\label{vpsloc-2}
 S_s(\textbf{G}) = \sum_{I=1}^{P_s} { e^{-i\textbf{G} \cdot \textbf{R}_I^s } } .
\end{equation}
In the last couple of equations, 
$\tilde{\rho} (\textbf{G}), \tilde{\rho}_\alpha (\textbf{G}), \tilde{\rho}_T (\textbf{G})$,
and $\tilde{v}_{PS}^{loc,s}(\textbf{G})$ are the coefficients of the Fourier expansions of the
corresponding real space functions, with
$\textbf{G}$ the reciprocal lattice vectors
($\tilde{f}(\textbf{G}) = 1/\Omega \int_\Omega f(\textbf{r})e^{-i\textbf{Gr}} d\textbf{r},
~f(\textbf{r}) = \sum_\textbf{G} \tilde{f}(\textbf{G}) e^{i\textbf{Gr}}$).
The complementary error function $erfc(x)$ arises from the
point-charges interactions screened by the Gaussian functions, with
$n$ an index running over cells in real space. The short-ranged nature of this
interaction ensures that the sum converges very fast: typically, only first
neighbours need to be considered.

\subsection{Total energy in the QM-MM implementation}

One of the key points in our hybrid approach is to conceive the MM atoms in the same way as
the pseudoions of the QM region within the PPW framework.
There are basically two
differences between MM and QM ions in this case: (i) the MM atoms do not
include a non-local pseudopotential term, and (ii) the MM ions can have a partial charge, which can
be either negative or positive, according to the charge parameter in the force field.
Hence, in our implementation, the electrostatic energy is extended to include the MM atoms:
\begin{equation}
 E_{es}[\rho] =  E_H[\rho] + E^{loc}_{PS}[\rho] + E_{em}[\rho] + E_{im}  
\label{qmmm-1}
\end{equation}
where $E_{em}[\rho]$ and $E_{im}$ represent, respectively, the interaction of the electron density with the
classical charges, and the Coulomb interaction between all
ions, both QM and MM. 
\begin{align}
E_{em}[\rho] &= \sum_{m=1}^{M} \sum_{l=1}^{P_m} \int{ \rho(\textbf{r}) 
{{ v_{MM}^{m} (|\textbf{r} - \textbf{R}_I|) d\textbf{r} }}  }  \\
E_{im} &= \dfrac{1}{2}  \sum_{I=1}^{T}{\sum_{J=1,J \neq I}^{T}{\dfrac{Z_IZ_J}{|\textbf{R}_I-\textbf{R}_J|}} } 
\end{align}
Here $M$ and $P_m$ are, respectively, the number of classical species and the number
of atoms for the $m$ species. The function $v_{MM}^m$ is the pseudopotential associated with the
classical species $m$, to be defined below.
\textbf{R}$_I$ is the position of every atom, irrespective of being quantum or classical,
and $Z_I$ is its charge, that will be typically a non-integer number in the MM region.
$T$ denotes the total number of atoms in the system ($T = \sum_s^{N} P_s + \sum_m^{M} M_m$).

The pseudopotential associated with the classical atoms, $v_{MM}^m$, has to verify a few properties:
has to be a smooth continuous function to be numerically tractable with Fast Fourier Transforms,
has to decay as the inverse of the distance $r$ at long ranges, and must avoid the divergence
when $r \rightarrow 0$.
We have adopted the functional form proposed by Laio et al. \cite{Laio2002} :
\begin{equation}
v_{MM}^m (|\textbf{r} - \textbf{R}_I|) = v_{MM}^m (r)=  Z_{m} \frac{r^4_{cm}-r^4}{r^5_{cm}-r^5}
\label{eq-vmm}
\end{equation}
with $m$ the classical atom species, $Z_{m}$ its charge, and $r_{cm}$ a
cutoff radius appropriate for every species. This function approaches $Z_{m}/r$ for $r \gg r_{cm}$, and
goes smoothly to $Z_{m}/r_{cm}$ for $r=0$. Even if the exact value of $v_{MM}$ at short ranges
is not critical, it has to be small enough not to become a trap for the electrons.
In the case of plane-wave basis, sharp MM potentials of positive species
may cause electronic charge localization on the
classical atoms: this is called the \textit{spill out} effect.
The possibility of electron density flowing to the MM region can be minimized using a classical
pseudopotential which varies softly and has a small magnitude at short distances.
The function defined in equation \ref{eq-vmm} satisfies these conditions, providing at the same time
an appropriate interaction between MM and QM atoms.

With these modifications, the electrostatic energy amounts to the following final form:
\[ E_{es}[\rho] = \dfrac{\Omega}{2} \sum_{\textbf{G}} \dfrac{4\pi}{G^2} \tilde{\rho}_T (\textbf{G}) \tilde{\rho}_T (-\textbf{G}) \]
\[+ \Omega \sum_{\textbf{G}} \left[ \sum_{s=1}^{N} S_s(\textbf{G}) \tilde{v}_{PS}^{loc,s}(\textbf{G}) 
+ \sum_{m=1}^{M} S_m(\textbf{G}) \tilde{v}_{MM}^m(\textbf{G})
-\dfrac{4\pi}{G^2}\tilde{\rho}_\alpha(\textbf{G}) \right] \tilde{\rho} (-\textbf{G}) \]
\begin{equation}\label{elec-final-qmmm}
+ \dfrac{1}{2} \sum_{I=1}^{T} \sum_{J \neq 1}^{T} Z_I Z_J \left[ \sum_{n=-n_{max}}^{n_{max}} \dfrac{erfc(|\textbf{R}_I + nL - \textbf{R}_J|\eta)}{|\textbf{R}_I + nL - \textbf{R}_J|} \right] 
 - \dfrac{\eta}{\sqrt{\pi}} \sum_{I=1}^{T} Z^2_I .
\end{equation}
This expression is identical to equation \ref{elec-final}, aside from the term involving the
structure factor $S_m(\textbf G)$ corresponding to the MM species, and from the fact that the
sums in the last two terms run over MM and QM atoms. The charge $\rho_\alpha$
now includes the contribution of the MM ions, and therefore it can take either negative
or positive values across space.

There is a non-electrostatic contribution to the QM-MM energy which is analogous to that
between atoms in the MM region:
\begin{equation}
 E_{LJ, im} = \sum_{s=1}^N \sum_{I=1}^{P_s} \sum_{m=1}^M \sum_{J=1}^{P_m} 
4 \epsilon_{sm} \left[ \left(\dfrac{\sigma_{sm}}{|\textbf{R}_{s,I} - \textbf{R}_{m,J}|}\right)^{12} 
- \left(\dfrac{\sigma_{sm}}{|\textbf{R}_{s,I} - \textbf{R}_{m,J}|}\right)^{6} \right]
\label{lennardqmmm}
\end{equation}
where now $\epsilon_{sm}$ and $\sigma_{sm}$ are the parameters for the Lennard-Jones
interaction of a classical atom of species $s$ with a quantum atom $m$.
This energy prevents the MM charges of negative sign from collapsing on the positive QM nuclei.
The MM subsystem in the present study involved H$_2$O molecules,
which were described through the SPC flexible water model (SPC/Fw) proposed
by Wu, Tepper and Voth \cite{flexiblespc}. The same set of parameters for $\sigma$ and $\epsilon$
were used in both the MM-MM and the QM-MM non-electrostatic interactions, given respectively
in equations \ref{lennardjones} and \ref{lennardqmmm}.

At this point it must be noticed that most force-fields, including the SPC/Fw potential,
do not consider any Coulomb interactions between atoms beloging to the same molecule.
In the present formulation, however, the electrostatic energy in
equation \ref{elec-final-qmmm} arises from pairwise interactions of every ion with all
the others, and those of intramolecular origin can not be easily individualized and
excluded from the rest.
A simple way to correct for this overcounting could be to separately compute the Coulomb
interactions inside each molecule, and then substract it from the total energy.
In this case, such a correction would be:
\begin{equation}
E_{intra} = \sum_{i=1}^{n_{H_2O}}  \dfrac{Z_O Z_H}{|\textbf{R}^i_O - \textbf{R}^i_{H1}|}
+ \dfrac{Z_O Z_H}{|\textbf{R}^i_O - \textbf{R}^i_{H2}|}
+ \dfrac{Z_H Z_H}{|\textbf{R}^i_{H1} - \textbf{R}^i_{H2}|}
\end{equation}
where $n_{H_2O}$ is the number of water molecules in the MM region, and $\textbf{R}^i_{O}$,
$\textbf{R}^i_{H1}$ and $\textbf{R}^i_{H2}$ are the positions of the three atoms belonging
to molecule $i$.

Finally, combining all the contributions together, we compute the total energy as:
\begin{equation}
 E_{tot}[\rho] = E_{es}[\rho] + T_e[\rho] + E_{XC}[\rho] + E_{PS}^{nl} + E_{LJ,im} + E_{LJ,MM} + E_{bond}
- E_{intra}
\label{total-energy-hybrid}
\end{equation}
\noindent where $E_{LJ,MM}$ considers the Lennard-Jones interactions within the MM region,
and $E_{bond}$ the intramolecular harmonic contributions between connected MM atoms (equation \ref{ebond}).

\subsection{Forces}

The atomic forces can be calculated for the QM and for the MM atoms
as the derivative of the total energy, equation \ref{total-energy-hybrid}, with respect to the ionic positions.
This leads to analytical forces in all cases.
For the QM atoms, there is no explicit dependence of $T_e[\rho]$ and $E_{XC}[\rho]$  on $R_I$,
and therefore only three terms survive in the derivative:

\begin{equation}
 F_I^{QM} = - \dfrac{dE_{tot}}{d \textbf{R}_I} =  - \dfrac {\partial E_{es}}{\partial \textbf{R}_I} - 
\dfrac {\partial E_{LJ,im}}{\partial \textbf{R}_I} - \dfrac {\partial E_{PS}^{nl}}{\partial \textbf{R}_I}
\label{force-qm}
\end{equation}

The former of these terms on the right hand side above can be developed as:
\[
 - \dfrac {\partial E_{es}}{\partial \textbf{R}_I} = \dfrac{Z_I}{2} \sum_{J\neq I}^{T} \sum_{n=-n_{max}}^{n_{max}} (\textbf{R}_I + nL - \textbf{R}_J)
 \times \left[ \dfrac{erfc(|\textbf{R}_I + nL - \textbf{R}_J|\eta)}{|\textbf{R}_I + nL - \textbf{R}_J|^3} 
+ \dfrac{\eta e^{-\eta^2|\textbf{R}_I + nL - \textbf{R}_J|^2}}{|\textbf{R}_I + nL - \textbf{R}_J|} \right]
\]
\begin{equation}
 + \Omega \sum_{G\neq0} i \textbf{G} e^{i\textbf{G}\cdotp\textbf{R}_I}
\left( \tilde{v}_{PS}^{loc,s}(\textbf{G})+\tilde{v}_{MM}^m(\textbf{G}) + 
\dfrac{4\pi Z_s}{G^2 \Omega} \right) \tilde{\rho}(-\textbf{G})
\label{forces-qmmm-1}
\end{equation}

On the other hand, the Lennard-Jones contribution
to the force is simply:

\begin{equation}
 - \dfrac {\partial E_{LJ,im}}{\partial \textbf{R}_I} = 
\sum_{m=1}^M \sum_{J=1}^{P_m} 4 \epsilon_{sm} \left[\dfrac{12\sigma_{sm}^{12}}
{|\textbf{R}_{s,I} - \textbf{R}_{m,J}|^{13}} 
-  \dfrac{6\sigma_{sm}^6}{|\textbf{R}_{s,I} - \textbf{R}_{m,J}|^{7}} \right]
\label{forces-qmmm-2}
\end{equation}

The contribution originating in the non local part of the pseudopotential energy is part of the
standard QM implementation and will not be discussed in the present context.

The expression for the forces on the atoms belonging to the MM subsystem will have two terms
in common with equation \ref{force-qm}, plus the pure MM contributions, whose derivatives
are straightforward:
\begin{equation}
 F_J^{MM} =  - \dfrac {\partial E_{es}}{\partial \textbf{R}_J} - 
\dfrac {\partial E_{LJ,im}}{\partial \textbf{R}_J} - \dfrac {\partial E_{LJ,MM}}{\partial \textbf{R}_J}
- \dfrac {\partial E_{bond}}{\partial \textbf{R}_J} + \dfrac {\partial E_{intra}}{\partial \textbf{R}_J}
\label{force-mm}
\end{equation}

\section{Assessment of the model: water dimer and aqueous liquid phase}

\subsection{The water dimer}

\begin{figure}
\begin{center}
\includegraphics[scale=0.46,keepaspectratio=true]{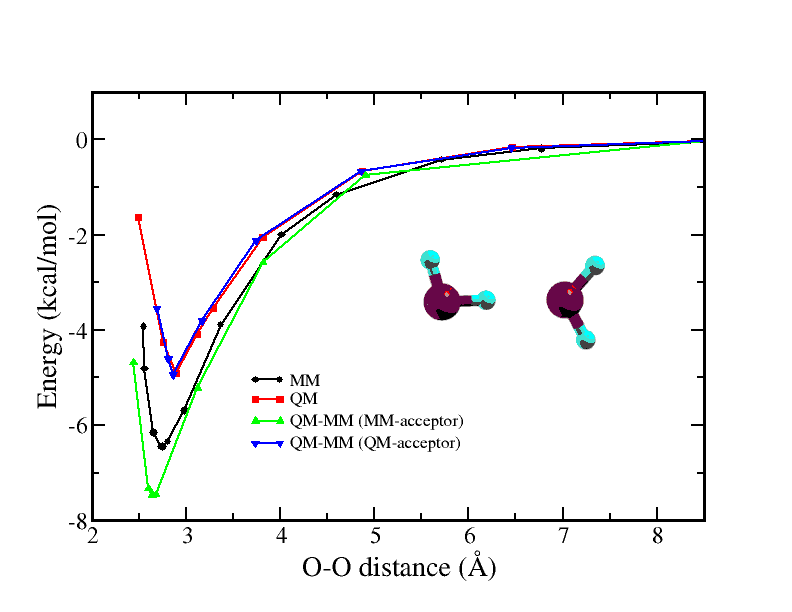}
\end{center}
\caption{Interaction energy for a water dimer as a function of the O-O distance,
according to DFT (QM), SPC/Fw (MM), and hybrid QM-MM calculations. In the dimer depicted in the inset,
the molecule on the right plays the role of H-bond acceptor.}
\label{water-dimer}
\end{figure}

The potential energy curve of a water dimer, with one  molecule in the QM domain
and the other in the MM region, was calculated as a first test.
The geometry of the dimer was optimized for a series of oxygen-oxygen separations.
It is important to note that there are two inequivalent configurations
in which this curve can be obtained, depending on whether the MM molecule plays the role
of donor or acceptor of the hydrogen bond. In the water dimer depicted in the inset of Figure \ref{water-dimer}, the hydrogen
bond donor is the molecule on the left and the acceptor is on the right.
Therefore, two curves can be obtained. 

Calculations were performed in a supercell of dimensions 36$\times$18$\times$18 bohr,
to minimize the interactions of the water dimer with their periodic images.
The cell dimension is longer along the $x$ axis because this
is the direction in which the potential energy is scanned.
The PBE approach to the DFT exchange-correlation energy \cite{pbe} in combination with
ultrasoft pseudopotentials \cite{vanderbilt1990} were adopted to compute total energies and forces.
The Kohn-Sham orbitals and charge density were expanded in planewaves up to a kinetic energy
cutoff of 50 and 200 Ry, respectively. An electronic mass of 400 a.u. was used to propagate
the wavefunctions according to the Car-Parrinello scheme.

Before discussing the QM-MM results, we will examine the curves provided by the pure
QM (DFT) and MM (SPC/Fw) methodologies. 
These are plotted in Figure \ref{water-dimer}. The difference in the description is readily
apparent: the QM curve is shifted to a weaker interaction and to a longer distance at
the minimum, with respect to the MM curve. Reported experimental values, 
of 2.98 \AA{}  for the oxygen-oxygen separation \cite{dyke1977,dyke1980},
and ranging from  3.6 to 5.2 kcal/mol for the magnitude of the interaction at the 
minimum energy geometry \cite{jcp_71_2703, nature_221_143, jquantspectroscradiattransf_25_83, molphys_74_639},
are better reproduced by the quantum-mechanical model, which provides
an optimized hydrogen-bond energy of 4.9 kcal/mol at 2.9 \AA{}.
Similar results have been obtained from previous DFT calculations based on both localized and
extended basis functions \cite{dixon1996,famulari1998, jp035869t, haynes2006,marzari2005}.
On the other hand, the classical-mechanics water dimer interaction, of nearly 7.4 kcal/mol,
is clearly above the available experimental data. The reason for this overestimation is that
the SPC and SPC/Fw potentials, as most water force-fields, are parameterized
to reflect the properties of the bulk liquid phase \cite{spc-e1987, flexiblespc}, where
the molecular dipole moments are significantly enhanced with respect to the isolated molecule (see next section).
As a matter of fact, the minimum of the MM curve, at nearly 2.7 \AA, is coincident with the first peak of the 
oxygen-oxygen radial distribution function (RDF) for liquid water at room temperature.

The QM-MM potential energy curves, presented in Figure \ref{water-dimer}
together with the results corresponding to the pure SPC/Fw and DFT calculations, turn out to be quite interesting.
As mentioned above, two configurations can be considered in this case,
depending on the identity of the donor of the H-bond.
Examination of the curves leads to the following observation: the QM-MM curve in which the acceptor
is the MM molecule, roughly reproduces the SPC/Fw curve, whereas the QM-MM curve where this role is
played by the QM molecule, is very close to the DFT results.
In other words, the QM-MM curves are essentially reflecting the identity of
the acceptor. This is a meaningful result, understandable when we recall
that most of the charge density involved in the
bond, and therefore the polarization effect, corresponds to the oxygen atom. The electron density
associated with the H atom is much lower and localized, and there is not a major effect if this electron
density is replaced by a bare pseudopotential.

Ideally, both QM-MM curves should be identical. In practice, however, a discrepancy is immanent
to all QM-MM models, since these curves are tied to the QM and MM Hamiltonians, which necessarily provide
different descriptions of the bond. For the calculations in the condensed phase, 
we expect that the difference between
the two kinds of interactions (involving the QM water molecule as the donor or as the acceptor)
will be substantially attenuated. The results in the next section, concerning the properties
in the bulk, suggest that this is certainly the case.

\subsection{Bulk phase properties}

Car-Parrinello molecular dynamics simulations were performed on a system
of 64 water molecules, of which one was described
quantum-mechanically and the rest classically. The simulations were conducted at 300 K
using the Nose-Hoover thermostat in a cubic box in periodic boundary conditions, 
with a density corresponding to 1 g/cm$^{3}$,
and a time-length of 6 ps.
A  planewaves basis of 25 Ry and the PW91 exchange-correlation functional were employed \cite{pw91}.

\begin{figure}
\begin{center}
\includegraphics[scale=0.46,keepaspectratio=true]{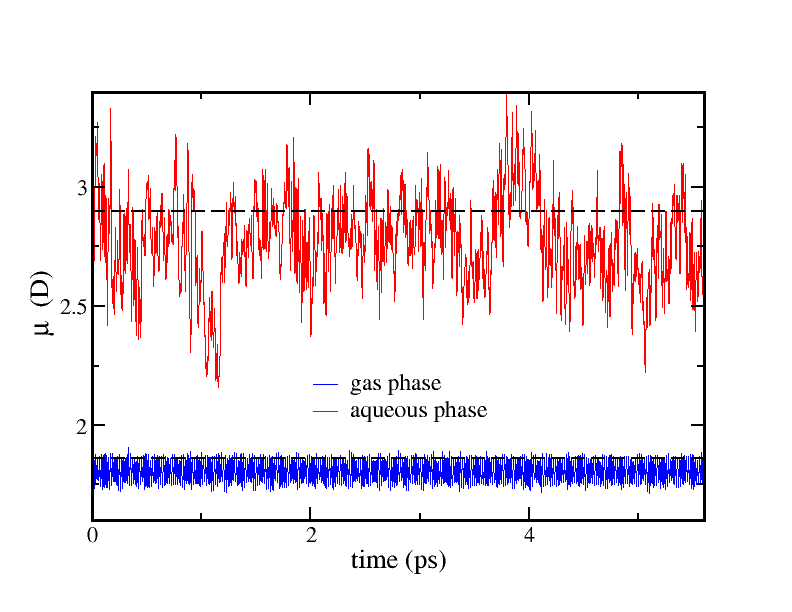}
\end{center}
\caption{Time evolution of the dipole moment of a quantum-mechanical water molecule in the gas
phase, and in an aqueous environment consisting of 63 classical molecules in a cubic cell.
The dashed horizontal lines
show the experimental values, from references \cite{dipole_gp1} and \cite{dipole_xr}.}
\label{dipolemoments}
\end{figure}

The average dipole moment for a water molecule in the gas phase obtained from the Car-Parrinello dynamics
is 1.81 D, very close to the experimental value of 1.86 D \cite{dipole_gp1, dipole_gp2}.
In the system of 64 H$_2$O molecules representing the bulk phase, the polarization exerted by the
classical environment raises the dipole moment of the QM molecule to an average of 2.88 D.
This number is in full agreement with reported estimates of 2.95 D from 
ab-initio simulations \cite{PhysRevLett.82.3308}, or 2.9$\pm$0.6 D from the x-ray 
structure factor \cite{dipole_xr}. Figure \ref{dipolemoments} depicts the computed dipole
moments of water as a function of time, both isolated and in the classical aqueous environment.
The dashed lines represent the experimental values. Such a good accord demonstrates that
the polarization effect is finely accomplished by the QM-MM scheme.

\begin{figure}
\begin{center}
\includegraphics[scale=0.46,keepaspectratio=true]{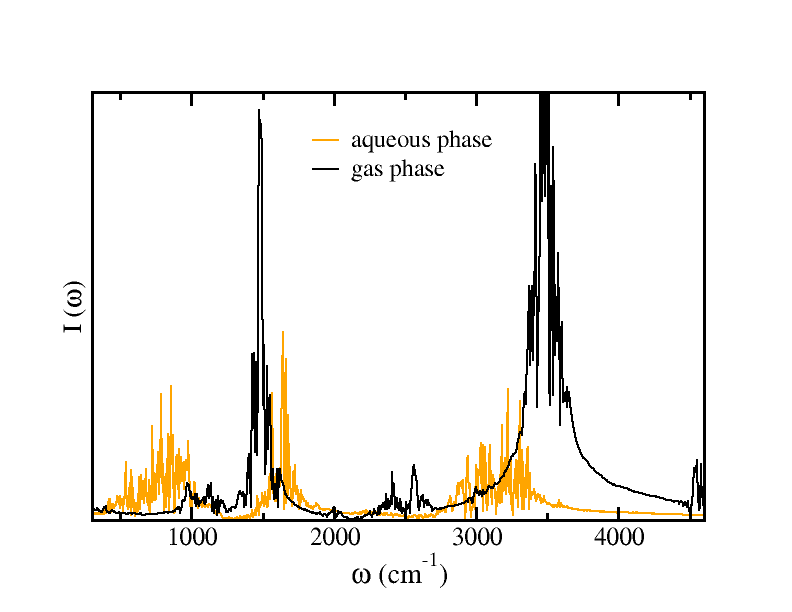}
\end{center}
\caption{Simulated vibrational spectra of water in the gas  and in
the liquid phases. In the later case, the aqueous environment is
represented by 63 classical molecules in  a cubic cell.
The computed frequency shifts are in qualitative agreement with IR spectroscopic data.}
\label{fig-wat-freqs}
\end{figure}

Figure \ref{fig-wat-freqs} presents the vibrational frequencies of the water molecule in
the gas and in the liquid phases, computed from the Fourier transform of the
time correlation function of the atomic velocities ${\bf v}(t)$ \cite{mcquarrie},
\begin{equation}
I(\omega) = \frac{1}{2\pi} \int_{-\infty}^{\infty} dt~ e^{-i\omega t} 
\frac{1}{N} \sum_{i=1}^N \langle {\bf v}_i(0) {\bf v}_i(t) \rangle. 
\label{acf-vel} \end{equation}
Depending on the system, on the vibrational mode,
and on the electronic mass,
ionic frequencies in Car-Parrinello dynamics may show  redshifts of a few percent  with respect
to spectroscopic data \cite{jcp_tangney}.
In the present case, the positions of the peaks fall between 100 to 200 cm$^{-10}$ below
the experimental frequencies, consistently with previous
Car-Parrinello simulations of H$_2$O \cite{jctc_2006, jpcb_119_8926}.
For the isolated molecule,
it is possible to recognize two groups of bands in Figure \ref{fig-wat-freqs}, 
corresponding to the stretching and bending modes,
centered at 3500 and 1480 cm$^{-1}$ respectively. In the liquid state these bands become
broader and noisier, with an additional set of peaks below 1000 cm$^{-1}$ arising from librations.
The first thing to note is that the stretching frequencies shift to lower wavenumbers
in the liquid, whereas the bending  experiences the opposite trend.
This is the same behavior as observed from IR spectroscopy, where the stretching in the liquid
is redshifted in about 300 cm$^{-1}$, and the bending mode is blueshifted in
nearly 50 cm$^{-1}$ \cite{water-spectra}. Our simulations in the liquid give
peaks which are spread and too much
splitted to establish unambiguously the magnitudes of these shifts; however, considering the center of mass
of the bands, it turns out that the shift of 300 cm$^{-1}$ in the stretching frequency
is pretty much reproduced by the QM-MM model, while the change in the bending frequency
appears overestimated by a factor of two.
These predictions for
the spectral shifts in water are, from a quantitative point of view, of a quality comparable
to that obtained from fully quantum-mechanical Car-Parrinello simulations \cite{jctc_2006,jpcb_119_8926}.

\begin{figure}
\begin{center}
\includegraphics[scale=0.46,keepaspectratio=true]{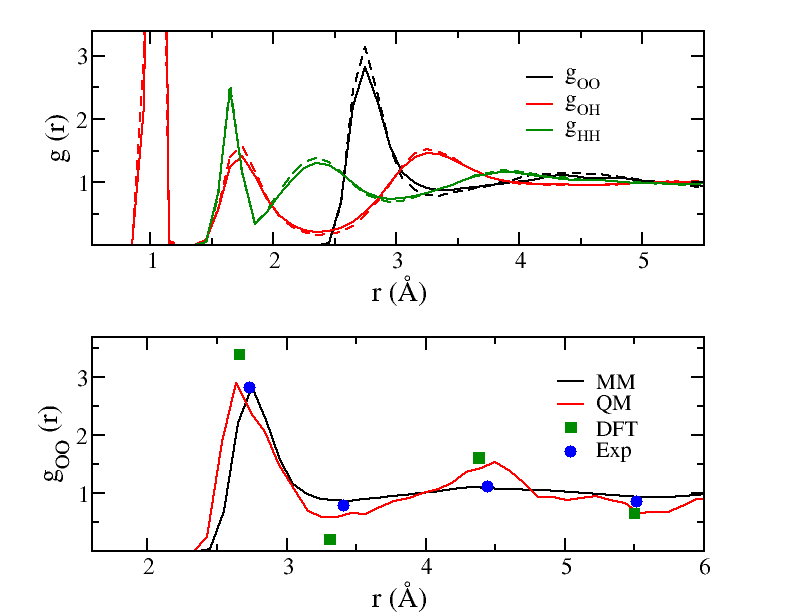}
\end{center}
\caption{{\bf Upper panel:} RDFs computed for all pairs
of MM atoms in the QM-MM system of 64 water molecules. The dashed lines correspond to the
curves obtained from classical molecular dynamics simulations with the SPC-Fw potential.
{\bf Lower panel:} RDFs for MM oxygen atoms (black) and
for the QM oxygen atom (red), from the simulations of the 64 molecules system.
The amount of data-points to construct the QM RDF is only 1/63 of that involved in
the MM RDF, which explains the uneven, unconverged structure of the former.
The  symbols show the positions of the maxima and the minima in the RDF obtained from
quantum-mechanical Car-Parrinello simulations (squares)
and from x-ray diffraction experiments (circles), extracted from references
\cite{marzari2005} and  \cite{als} respectively.}
\label{fig-gdr}
\end{figure}

Radial distribution functions for the model of 64 water molecules are displayed in Figure \ref{fig-gdr}.
Classical molecular dynamics simulations with the SPC-Fw potential in PBC
were performed in the same system with the LAMMPS code \cite{lammps}.
The upper panel confirms that the electrostatic description of the MM atoms within
the QM-MM approach, including the correction to the intramolecular Coulomb forces,
reproduces the dynamics dictated by the SPC-Fw force-field. 
The subtle discrepancies between the classical and the QM-MM curves are attributable to differences
in the length of the simulations, to the distinctive numerical implementations,
and eventually to the presence of the QM water molecule. The radial distribution
function corresponding to the quantum-mechanical oxygen atom is presented in the
lower panel of Figure \ref{fig-gdr}. In comparison to SPC-Fw, this curve is more structured,
with maxima and minima appearing respectively above and below. Its shape does not seem to
be converged, probably because insufficient sampling: note that in this case the RDF is built from
a single water molecule out of 64. Various studies of liquid water using the Car-Parrinello 
method with GGA functionals and an electronic mass comparable to the one employed here,
have found overstructured oxygen-oxygen RDFs at room temperature and
pressure \cite{schwegler-I,schwegler-II,marzari2005,jpcb_119_8926},
suggesting that in these conditions this approach represents bulk water in a glassy 
or supercooled state. Data from one of these works is depicted
in Figure \ref{fig-gdr}. The radial distribution function of the QM oxygen atom in our QM-MM
simulation turns out to be intermediate between those obtained from pure 
classical and pure quantum-mechanical Car-Parrinello molecular dynamics.

\subsection{Considerations on computational efficiency}

\begin{figure}
\begin{center}
\includegraphics[scale=0.46,keepaspectratio=true]{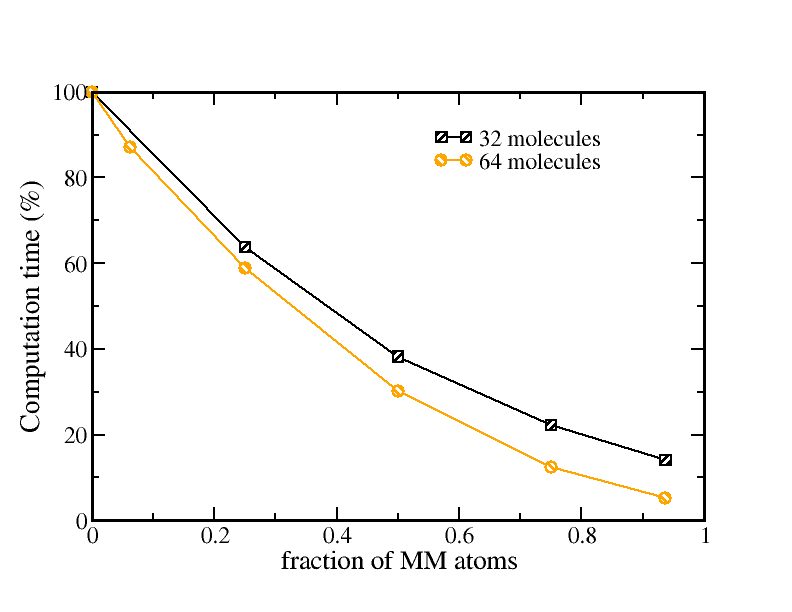}
\end{center}
\caption{Dependence of the total computational time with the size of the classical
domain, keeping constant the total (QM + MM) number of atoms.
The benchmarks correspond to 100 Car-Parrinello steps parallelized on 4 processors for
systems of 64 and 32 water molecules in a cubic unit cell. Tests on different number of processors,
up to 16, provide identical trends.}
\label{comp_time}
\end{figure}

At variance with other QM-MM schemes in the PPW framework, in which the classical
region does not have to be included in the simulation box, in the present treatment the
MM atoms need to be contained in the supercell together with the QM atoms.
Then, the amount of planewaves and the size of real space grids are the same or about
the same as in a quantum-mechanical calculation with an equal number of total atoms,
and therefore the QM-MM implementation does not involve any significant decrease in memory requirements.
In spite of this, the reduction of the QM region cuts the quantity of Kohn-Sham states to be
evolved in the Car-Parrinello dynamics,
which may have a substantial impact on the overall computing time.
As a matter of fact, the computational effort in the quantum-mechanical Car-Parrinello
scheme for a given unit cell size is approximately proportional to the number of atoms,
with a slope larger than 1, which tends to increase with system dimensions.
As a consequence, the speedup achieved by the QM-MM approach is roughly linear with
the replacement of QM by MM atoms. This behavior is reflected in
Figure \ref{comp_time}, which illustrates the relative decrease in computation time
with the increase of the number of atoms represented classically, for the cases
of 32 and 64 water molecules. As expected,
the gain in performance becomes more significant as the system grows bigger.

In ab-initio modelling of solid interfaces in contact with a bulk liquid, 
the solvent fills a major fraction 
of the supercell, often representing between 1/2 to 2/3 of the total atoms.
In the solid-liquid interface model for anatase (101) discussed in the next section,
for instance,
the number of classical atoms is 144, out of a total of 264. In this situation, the
QM-MM calculation turns out to be 5 times faster than a full QM simulation.
In the case of a quantum-mechanical water molecule surrounded by  63 classical ones,
examined in the previous section, the acceleration goes above one order of magnitude.

\section{Water vibrational frequencies at the solid-liquid interface of TiO$_2$ anatase}

Many of the most relevant applications of TiO$_2$ implicate the solid-liquid interface,
which entails a serious challenge to first-principles modelling, given the need
for extensive simulations to achieve an appropriate configurational sampling of the fluid phase.
Thus, ab-initio molecular dynamics investigations
considering the water liquid phase in contact with titania
have been carried out only in a limited number of occasions,
to address the rutile (110) \cite{prb_82_161415, sprik_jctc_2010, sprik2010-2, sprik2014}
and the anatase (101) and (001) surfaces \cite{sumita}.
Instead, there is a large number of DFT studies which have examined 
stoichiometric and defective titania interfaces in the presence of just a few water monolayers, typically
ranging from one to three \cite{lindan2003, selloni2004, selloni2004-2,HarrisQ2004, 
Lindan2005, bonapasta2008, prl_80_762, jpcc_116_9114, english1}.
In this section we illustrate the applicability of our hybrid quantum-mechanics molecular-mechanics
scheme through the calculation of the water vibrational frequencies at the anatase (101)
interface. We aim at determining to what extent the explicit inclusion of the liquid phase
affects the dynamical properties of the first adsorbed layers. To this end,
results from QM-MM molecular dynamics simulations representing the liquid phase, are
compared with those coming from QM simulations
incorporating just one or two H$_2$O monolayers.

\begin{figure}
\begin{center}
\includegraphics[scale=0.38,keepaspectratio=true]{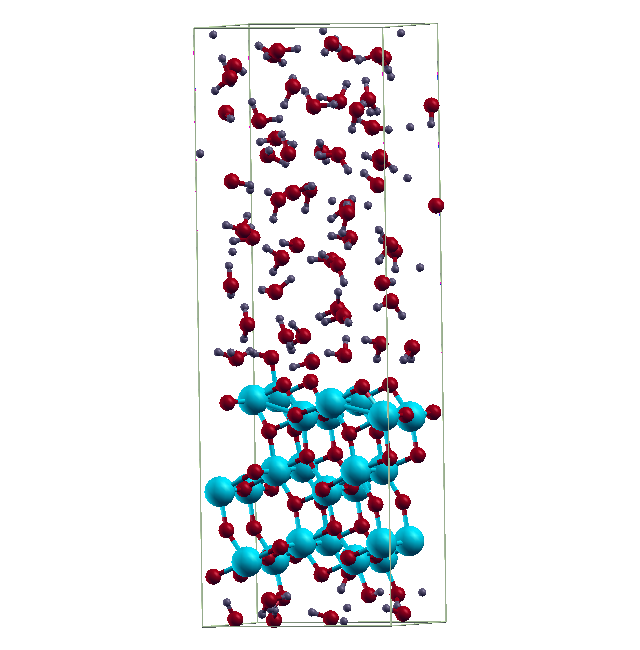}
\end{center}
\caption{Slab model used in the QM-MM simulations of the TiO$_2$ anatase (101) surface 
in contact with a bulk water phase. The  water molecules
were represented classically, excepting those forming the first and second adsorbed layers.}
\label{anatase-model}
\end{figure}

The DFT parameters concerning planewave basis, pseudopotentials, and exchange correlation functional,
were the same as employed to describe the water bulk phase in the previous section.
The anatase (101) surface was represented using a 2$\times$2 supercell,
containing six layers of TiO$_2$ units. Cell dimensions were
7.56$\times$10.24$\times$22.67 \AA$^3$ for the QM calculations. In the QM-MM simulations 
the $z$-parameter was extended to 30.23 \AA~to accomodate the aqueous phase,
consisting of 16 QM plus 48 MM water molecules. In particular,
the first two water monolayers adjacent to the solid surfaces were modelled
quantum-mechanically, to get the corresponding vibrational frequencies and
to avoid a direct interaction between SPC-Fw water and titania.
Molecular dynamics simulations were performed at 300 K with the Nose-Hoover thermostat,
with sampling windows of 6 ps.
Figure \ref{anatase-model} displays the model structure employed in the QM-MM calculations
of the solid-liquid interface.

\begin{figure}
\begin{center}
\includegraphics[scale=0.46,keepaspectratio=true]{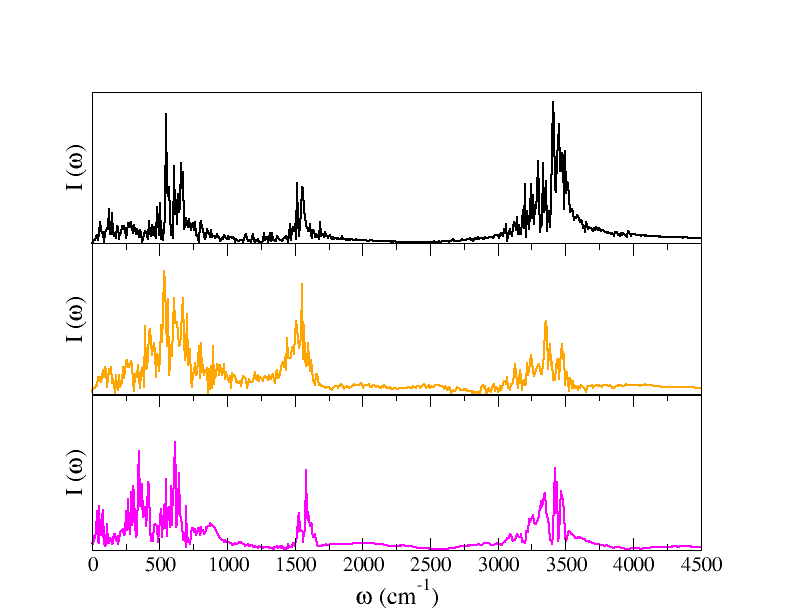}
\end{center}
\caption{Simulated vibrational spectra corresponding to the first water monolayer adsorbed
on the anatase (101) surface, at different coverages: monolayer (top panel),
bilayer (middle panel), and bulk liquid (bottom panel).}
\label{anat-1st}
\end{figure}

Figure \ref{anat-1st} shows the vibrational density of states corresponding to
the water molecules directly adsorbed on the
TiO$_2$ surface, computed through equation \ref{acf-vel}. The three panels compare three different
coverages: monolayer (top), bilayer (center), and the liquid environment (bottom).
For the single layer, the stretching mode, appearing at around 3500 cm$^{-1}$, is
the strongest one, vaguely resembling the vibrational patterns of water in the gas phase.
With the incorporation of a second layer, the intensities arising from librations and bending
become larger than that associated with the stretching, which in turn moves slightly to the left.
This trend is similarly observed in the presence of the bulk liquid, where the librations dominate the
spectra.
Hence, in general terms it is observed that the vibrational behavior of the H$_2$O molecules in
the first layer, roughly shifts from gas-like to liquid-like as the degree of hydration
is increased. The main features in our computed spectra are in line with those
obtained from neutron scattering experiments on water confined in anatase nanoparticles,
where librations and bending are predominant, and the stretching absorption band
is shifted to lower frequencies with respect to the gas phase \cite{jpca_111_12584}.

\begin{figure}
\begin{center}
\includegraphics[scale=0.46,keepaspectratio=true]{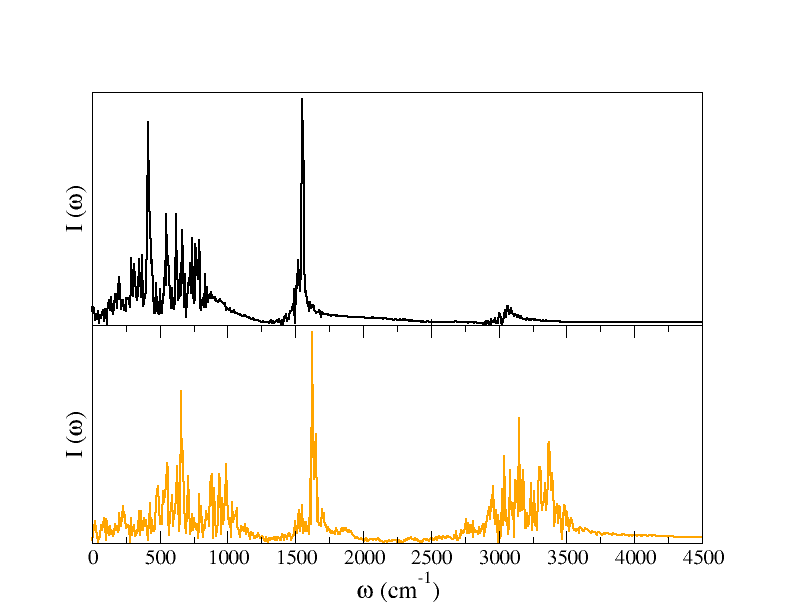}
\end{center}
\caption{Simulated vibrational spectra corresponding to the second water monolayer adsorbed
on the anatase (101) surface, for a bilayer (top panel) and 
the bulk liquid phase (bottom panel).}
\label{anat-2nd}
\end{figure}

The aqueous media has a major influence on the spectra of the second water layer.
Figure \ref{anat-2nd} shows that in the absence of the liquid environment,
the stretching of the water molecules in the second layer is very weak 
in comparison with librations and bending. The typical bulk-water features are recovered as
the MM environment is included in the simulation.

The above results suggest that the internal degrees of freedom
of a water molecule in a single layer on anatase (101)
retain some of the character they have in the gas phase.
Moreover, the sole inclusion of a second H$_2$O layer appears to be enough
to recreate, at least to some extent, the dynamics  in the presence
of a bulk aqueous phase. This behavior can be tracked to the hydrogen bond network
arising in each case. For a single adsorbed layer, the H$_2$O molecules are tightly bound to the
five-coordinate Ti atoms, with one or both of their hydrogen atoms forming relatively
weak hydrogen-bonds with the bridging oxygen sites on the surface. On the other hand, an inspection of the
trajectories in the presence of a second row of solvent molecules reveals that, most
of the time, the molecules in the first layer are involved in hydrogen-bonds with at least another
H$_2$O molecule, either as a donor or as an acceptor.  In the liquid phase the number
of hydrogen bonds that can be formed, and the polarization effects, are even larger. Yet, the
incorporation of a few water molecules beyond the first monolayer is sufficient
to induce on the adsorbate, a dynamical behavior very close to the one corresponding to
full hydration.

\section{Summary}

We have presented an approach to perform hybrid quantum-mechanics molecular-mechanics
simulations with the Car-Parrinello method in the context of the pseudopotential-planewaves
setting. At variance with other QM-MM implementations existing in planewaves codes,
in the present approach the classical atoms are treated on the same footing
as the quantum-mechanical ions, which naturally leads 
to periodic boundary conditions for the totality of the system.
Thus, all QM and MM atoms need to be contained within the same real space
grid determined by the simulation cell.
As a consequence, the size of the MM region has an impact on the computational cost,
and this method would not be convenient for extended systems where this region is
extremely large, or exceeds by far the size of the QM part.
It turns out that, for a given unit cell, the scalability is approximately linear with the substitution
of QM by MM atoms. In typical calculations of solid-liquid interface models involving a few
hundred atoms, in which
the solvent represents more than one half of the system, speedup
factors above five can be attained with the present scheme.
We applied our implementation to the computation of the vibrational spectra of water
adsorbed at the TiO$_2$ anatase (101) surface, at various coverages. 
It was found that the presence of a second monolayer of water molecules
is enough to mimick the effect of an aqueous environment on the vibrational frequencies
of the first adsorbed layer.
This methodology seems particularly suited for molecular dynamics simulations in condensed matter
systems including one or more fluid phases. Solutions, nanoconfined fluids,
or solid-liquid and liquid-liquid interfaces,
are all examples where this scheme could be extremely valuable.

\section{Acknowledgments}

We express our gratitude to Ivan Girotto and Paolo Gianozzi for precious help
related to the structure and parallelization of the Quantum-Espresso code.
We also thank Davide Ceresoli for useful discussions.
This study has been partially supported by grants
of the Agencia Nacional de Promocion Cientifica y
Tecnologica de Argentina, PICT 2012-2292, and UBACYT 20020120100333BA.
We acknowledge CSC-CONICET for granting the use of
the TUPAC HPC cluster, which allowed to perform most of the computations
included in this work.

\vspace{2cm}

{\bf Corresponding author's e-mail:} damian@qi.fcen.uba.ar

\end{document}